\documentclass[aps,showpacs,prx,twocolumn,superscriptaddress, floatfix]{revtex4-2}

\usepackage[colorlinks,urlcolor=blue,linkcolor=blue,anchorcolor=blue,citecolor=blue]{hyperref}

\usepackage{amsmath}
\usepackage{graphicx}
\usepackage{subfigure}
\usepackage{color}
\usepackage{amsfonts}
\usepackage{tikz}
\usepackage{esint}
\usepackage{mathrsfs}
\usepackage{mathptmx}
\usepackage{lettrine}
\usepackage{lipsum}

\begin{document}
\title{Reweight-annealing method for evaluating the partition function via quantum Monte Carlo calculations}

\author{Yi-Ming Ding}
\affiliation{State Key Laboratory of Surface Physics and Department of Physics, Fudan University, Shanghai 200438, China}
\affiliation{Department of Physics, School of Science and Research Center for Industries of the Future, Westlake University, Hangzhou 310030,  China}
\affiliation{Institute of Natural Sciences, Westlake Institute for Advanced Study, Hangzhou 310024, China}

\author{Jun-Song Sun}
\affiliation{School of Physics, Beihang University, Beijing 100191, China}

\author{Nvsen Ma}
\affiliation{School of Physics, Beihang University, Beijing 100191, China}

\author{Gaopei Pan}
\affiliation{Institut für Theoretische Physik und Astrophysik and Würzburg-Dresden Cluster of Excellence ct.qmat,
	Universität Würzburg, 97074 Würzburg, Germany}

\author{Chen Cheng}
\affiliation{Key Laboratory of Quantum Theory and Applications of MoE, Lanzhou Center for Theoretical Physics, and Key Laboratory of Theoretical Physics of Gansu Province, Lanzhou University, Lanzhou, Gansu 730000, China}

\author{Zheng Yan}
\email{zhengyan@westlake.edu.cn}
\affiliation{Department of Physics, School of Science and Research Center for Industries of the Future, Westlake University, Hangzhou 310030,  China}
\affiliation{Institute of Natural Sciences, Westlake Institute for Advanced Study, Hangzhou 310024, China}

\begin{abstract}
Efficient and accurate algorithm for partition function, free energy and thermal entropy calculations is of great significance in statistical physics and quantum many-body physics. Here we present an unbiased but low-technical-barrier algorithm within the quantum Monte Carlo framework, which has exceptionally high accuracy and no systemic error. Compared with the conventional specific heat integral method and Wang-Landau sampling algorithm, our method can obtain a much more accurate result of the sub-leading coefficient of the entropy consistent with theory. This method can be widely used in both classical and quantum Monte Carlo simulations and is easy to be parallelized on computer. 
\end{abstract}

\maketitle
% ---------------------------------------------------------
% ---------------------------------------------------------
% ---------------------------------------------------------
% ---------------------------------------------------------
% ---------------------------------------------------------
\section{Introduction}
Quantum Monte Carlo (QMC) is one of the powerful tools in calculating large-scale and high-dimensional quantum many-body systems~\cite{Sandvik1999,Sandvik2010,sandvik2019,Syljuasen2002,yan2019sweeping,suzuki1977monte,PhysRevB.26.5033,suzuki1976relationship,PhysRevE.66.066110,huang2020worm,fan2023clock,yan2020improved}. Although many observables, e.g., energy and specific heat, can be obtained from QMC through sampling the partition function (PF), the value of the PF itself remains difficult to calculate directly. To tackle this problem, the specific heat integral ($C$-int) method~\cite{emonts2018monte} and Wang-Landau (WL) algorithm~\cite{PhysRevLett.86.2050, PhysRevLett.90.120201} have been widely adopted to calculate PF and related quantities, such as free energy and entropy. The former method crudely calculates the numerical integration of $S(T)=S(\infty)-\int_T^{\infty} {C(T')}/{T'}\mathrm{d}T'$ to obtain the entropy $S(T)$~\footnote{$T$ means the temperature and $C$ is the specific heat. Entropy at infinite temperature $S(\infty)$ is known because each configuration has equal weight.}, after which we can calculate PF indirectly. However, this method usually requires a plethora of computational resources to lower the systemic error especially when the specific heat has sharp variation and huge fluctuation. The key idea of the latter method, on the other hand, is to achieve the (effective) density of states of the PF by initially allocating them some default values, then optimizing them to the real ones by flattening some histograms \cite{PhysRevLett.86.2050, PhysRevLett.90.120201, cond-mat/9610041}. This method suffers from the problem of error saturation, which prevents us from achieving a result with arbitrary precision, and its convergence is not generally controllable \cite{10.1063/1.2803061, PhysRevE.72.025701}. 

On the other hand, the high-precision PF and its related quantities, such as free energy and entropy, are significantly important in numerical calculations since it means more effective information.
For phase transitions no matter within or beyond the Ginzburg-Landau paradigm, an efficient and accurate way to straightforwardly obtain PF would be crucial, because its (high order) derivative reveals potential phase transitions.  The thermal entropy also relates to the distinctiveness of many novel condensed matters, such as spin ice~\cite{snyder2001spin,skjaervo2020advances,udagawa2021spin,bramwell2020history,udagawa2021spin,CJHuang2018}, topological order~\cite{jiang2012identifying,castelnovo2012spin,sachdev2018topological,levin2006detecting,yan2021topological,yan2022triangular}, fracton phases~\cite{pretko2020fracton,nandkishore2019fractons,gromov2024colloquium,zhou2022evolution}, etc. 
Besides, the precision of entropy can even reveal important intrinsic physics. In conformal field theory (CFT), the prefactor in front of the sub-leading term of entropy often carries universal information~\cite{Calabrese2008entangle,Fradkin2006entangle,Nussinov2006,Nussinov2009,CASINI2007,JiPRR2019,ji2019categorical,kong2020algebraic,XCWu2020,JRZhao2020,XCWu2021}. Even a small numerical error may make our concerned quantities deviate a lot because the sub-leading term is much more sensitive than the leading one. 

In this paper, we present a more accurate and efficient method for calculating PF, dubbed reweight-annealing algorithm, within the framework of Monte Carlo (MC) simulation, either quantum or classical.
It is inspired by thermal and quantum annealing algorithm in the quantum simulation area~\cite{Huse1986,Car2002,Lucas2014,kadowaki1998quantum,Car2002,Montanari2009,Heim2015,yan2023emergent,yan2023quantum,ding2023quantum} and recent developments of the incremental trick in the entanglement calculations of QMC~\cite{PhysRevE.95.062132,d2020entanglement,zhao2022measuring,zhao2022scaling,pan2023stable,zhou2024incremental}, while the earliest idea can date back to the thermodynamic integration~\cite{Frenkel1986thmIntUsed,neal1993probabilistic,XLMeng1998normalizing}, and many of its variants found the applications in both condensed matter physics and high energy physics~\cite{Forcrand2001snakeAlgorithm,Forcrand2005thmInt,Forcrand2005thmIntHooftLoop,MCaselle2003thmIntExp,LPollet2008thmIntQMC,Hasenbusch2009thmIntCasimir,RMelko2010thmIntMutInfo}. 
The method not only has high accuracy for calculating PF, but is unbiased, easy to implement and naturally can be parallelized on computer. With the annealing scheme we introduce below, a polynomial complexity can be proved, enabling better control over the computational precision for large systems.

This paper is arranged as follows. In Sec.~\ref{eq:method}, we elaborate the algorithmic and technical details of our method, including the annealing scheme, then to demonstrate the efficacy of the method, we provide three examples in Sec.~\ref{sec:examples}. In the end, a summary and discussions are put in Sec.~\ref{eq:summary}.

% ---------------------------------------------------------
% ---------------------------------------------------------
% ---------------------------------------------------------
% ---------------------------------------------------------
% ---------------------------------------------------------
\section{Method}\label{eq:method}
\subsection{Basic idea}
\begin{figure}[!t]
\includegraphics[width=\columnwidth]{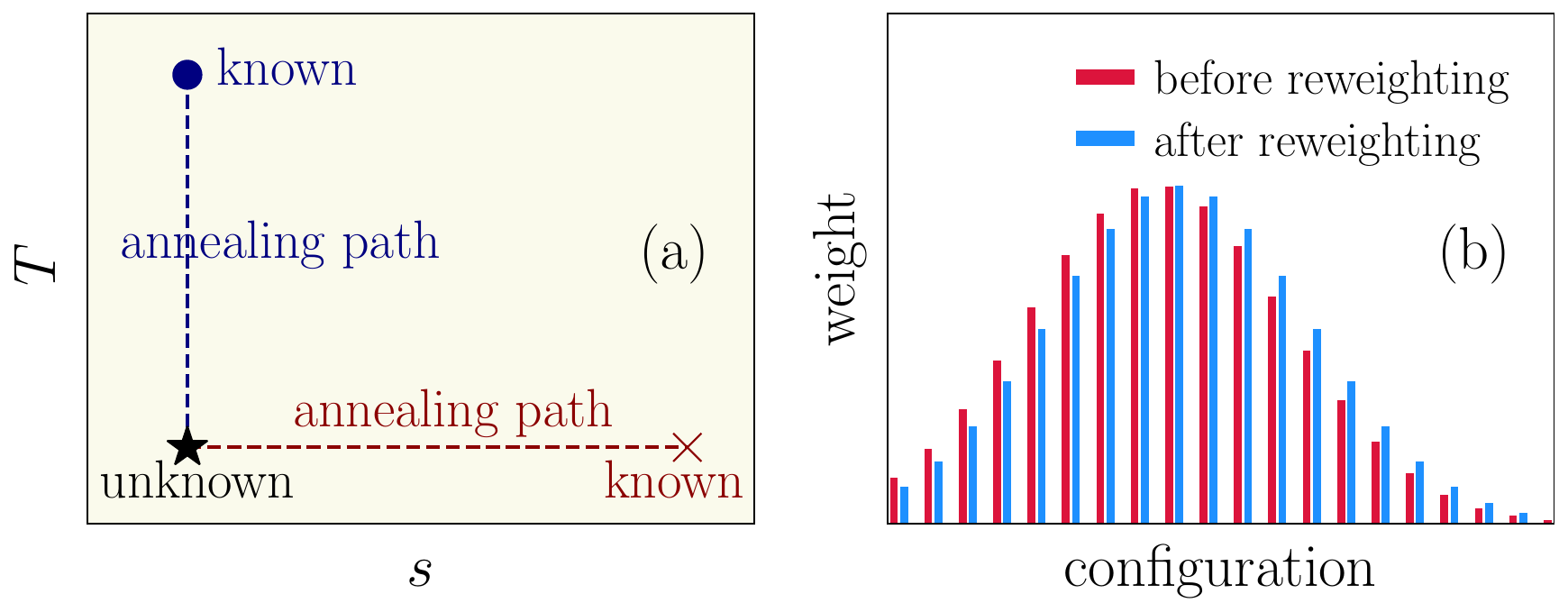}
\caption{Schematic diagram of the reweight-annealing method. (a) Finding a reference point and the annealing path connecting the unknown point to it in the parameter space. $T$ and $s$ correspond to the thermal and quantum version, respectively; (b) Using a sampled distribution (red) to represent another distribution (blue) through reweighting a same configuration. If these two distributions are close to each other, the effect of the reweighting would be good since the importance sampling can approximately be kept.}
\label{fig:idea}
\end{figure}

Given a partition function $Z(\vec{p})$, where $\vec{p}=[p_1,p_2,\cdots]$ denotes the parameters in the PF such as the temperature $T$ and some Hamiltonian parameter $s$. For example, Fig.\ref{fig:idea} (a) shows a diagram of the two parameters.
For two different parameter points $\vec{p}'$ and $\vec{p}''$, we define the reweighting ratio which can be measured in Monte Carlo 
\begin{equation}\label{eq:ratio}
    \frac{Z(\vec{p}')}{Z(\vec{p}'')}=\bigg\langle \frac{W(\vec{p}')}{W(\vec{p}'')} \bigg\rangle  _{\vec{p}''}
\end{equation}
%Finding an estimator for $r(\vec{p}',\vec{p}'')$ is the first step in our method, which generally depends on the model we investigate and the MC framework we adopt. It can usually be expressed as a form like $\big\langle W(\vec{p}')/W(\vec{p}'') \big\rangle  _{\vec{p}''\vec{p}'',\text{MC}}$, 
where $W(\vec{p}')$ and $W(\vec{p}'')$ denote the weight of the same configuration under different parameters $\vec{p}'$ and $\vec{p}''$, and the subscript indicates the result comes from the MC simulation under parameter $\vec{p}''$. %For simplicity, we will omit the text $\text{MC}$ henceforth.
%In this way, $r(\vec{p}',\vec{p}'')$ can be simulated via MC in principle.
As an example, we take $\vec{p}=\beta$, where $\beta$ denotes the inverse temperature. Under the framework of stochastic series expansion (SSE)\cite{Sandvik1999,Sandvik2010}, we can then derive $Z(\beta')/Z(\beta'')=\langle  (\beta'/\beta'')^n \rangle_{\beta''}$, where $n$ is the number of non-identity operators in the SSE simulation. The details of the $Z(\beta')/Z(\beta'')$ can be found in Appx.~\ref{appdx:estimators}.

The second step is to identify a reference point $\vec{p}_0$, of which the value of $Z(\vec{p}_0)$ has already been known. Therefore, by evaluating $Z(\vec{p}_0)/Z(\vec{p}'')$, we can directly determine the value of $Z(\vec{p}'')$. In the example that $\vec{p}=\beta$, apparently $Z(0)=d^N$ is pre-knowledge because all the configurations are equally important at infinite high temperature, where $N$ and $d$ are the number of particles and the degree of freedom of a single particle respectively. 

However, when $\vec{p}_0$ and $\vec{p}''$ are far away in the parameter space, the value of $Z(\vec{p}_0)/Z(\vec{p}'')$ would be close to zero or infinity, resulting in the problem of inefficiency and low precision for the MC simulation. This can be easily understood in the reweighting frame as shown in Fig. \ref{fig:idea} (b). If we want to use a well-known distribution $Z(\vec{p}'')=\sum W(\vec{p}'')$ to calculate another distribution $Z(\vec{p}')=\sum W(\vec{p}')$ via resetting the weight of samplings, the weight before/after the resetting of the same configuration should be close to each other. It is still an importance sampling in this sense and requires that $\vec{p}'$ and $\vec{p}''$ should be close enough \cite{neal1993probabilistic, neal2001annealed}.

Based on this consideration, practically, we divide the calculation of of Eq. (\ref{eq:ratio}) into many steps to enhance precision, i.e. 
\begin{equation}\label{eq:divide}
    \frac{Z(\vec{p}_0)}{Z(\vec{p}'')}=\prod_{k=1}^m \frac{Z(\vec{p}_{k-1})}{Z(\vec{p}_k)}
\end{equation}
where $\vec{p}_{m}\equiv \vec{p}''$ and $Z(\vec{p}_{k-1})/Z( \vec{p_{k}})\to 1$ as long as $\vec{p}_{k}\to \vec{p}_{k-1}$. Here, $\vec{p}$ can be perceived to gradually change from the point of interest $\vec{p}''$ to the reference point $\vec{p}_0$ along a trajectory (annealing path) in the parameter space, akin to an adiabatic process or annealing as shown in Fig.\ref{fig:idea} (a). 
Therefore by calculating Eq.~(\ref{eq:divide}), not only we are able to achieve a higher precision of $Z(\vec{p}'')$, but concurrently track the variation of PF along this annealing path. This allows us to further analyze the variation of the free energy (and its derivatives) of the system.

% Apparently, the annealing path that $\vec{p}$ follows determines how large $m$ should we need and how $\vec{p}$ varies to perform the calculations of Eq.~(\ref{eq:divide}). It can be expected that, similar to conventional thermal annealing or quantum annealing, should the annealing trajectory approach a critical point, an increased number of divisions would be necessary in its vicinity. 
% However, the pseudo-automatic annealing process which we will introduce later is not only affected little by that, but economic and makes the algorithm have a polynomial complexity with the system size. 

% ---------------------------------------------------------
% ---------------------------------------------------------
% ---------------------------------------------------------
% ---------------------------------------------------------
% ---------------------------------------------------------
\subsection{Thermal and quantum reweight-annealing}
As it shown in Fig.~\ref{fig:idea}, if $\vec{p}=\beta$, as previously mentioned, the annealing path connects an unknown inverse temperature point $\beta''$ to the infinite temperature point $\beta_0$, which serves as the reference. For convenience, we call this kind of process the thermal reweight-annealing (T-Re-An). 
Similarly, we can consider the so-called quantum reweight-annealing (Q-Re-An) by setting $\vec{p}=s$, where $s$ comes from $H(s) = sH_0 + (1-s)H_1$, which is the archetypal Hamiltonian of quantum annealing with $Z(s)=e^{-\beta H(s)}$. Here, $H_1$ correspond to our reference point $s=0$, and we would like to obtain the PF related to $H_0$. Different from the thermal version, the choice of $H_1$ here is much more flexible. For instance, we can take $H_1=\sum_i \sigma_i^x$. Then at the zero temperature ($\beta\to\infty$), $H_1$ only has a unique ground state, thus its thermal entropy is minimized to be zero, resulting in $\ln Z(0) = -\lim_{\beta\to\infty}\beta E$, where $E=-N$ is the ground state energy of $H_1$. More details about the T-Re-An and Q-Re-An scheme can be found in the SM.

% ---------------------------------------------------------
% ---------------------------------------------------------
% ---------------------------------------------------------
% ---------------------------------------------------------
% ---------------------------------------------------------
\subsection{Automatic annealing process}
Here we provide a scheme to automatically regulate the alteration of $\vec{p}$ throughout the annealing process.
This not only serves as an economic way for the division in Eq.~(\ref{eq:divide}), but can avoid the problem of precision loss when different $Z(\vec{p}_{k-1})/Z(\vec{p}_k)$ differ by several orders of magnitude in the computer. For demonstration, we discuss the thermal case, i.e. $\vec{p}=\beta$ in this section, and the case that $\vec{p}=s$ is similar.

For $\beta_0<\beta$, when $\beta_0$ and $\beta$ are far away in the parameter space, to ensure importance sampling, we divide the interval $[\beta_0,\beta]$ into $m$ subintervals $[\beta_0,\beta_1]$, $[\beta_1,\beta_2]$, $\cdots$, $[\beta_{m-1},\beta_m]$, where $\beta_m\equiv \beta$, i.e.
\begin{equation}\label{eq_appx:divide}
    \frac{Z(\beta_0)}{Z(\beta)}=\prod_{k=1}^m \frac{Z(\beta_{k-1})}{Z(\beta_k)}
\end{equation}
as Eq.~(\ref{eq:divide}). Each subinterval $[\beta_{k-1}, \beta_k]$ ought to be sufficiently narrow so that the distribution at $\beta_{k}$ can well samples the configurations at $\beta_{k-1}$. Otherwise, the sampling process would be inefficient. Then the natural questions following are how to partition $[\beta_0,\beta]$ and the how many subintervals we need for Eq.~(\ref{eq_appx:divide}).

One crude way is just dividing $[\beta_0,\beta]$ into equal subintervals or in a geometric progression. As long as the division number is large enough, the result would be accurate. However, this way is not economic. We can expect that when $\beta\to 0$ (approaching infinite high temperature), the division can be more sparse compared to that at low temperatures. 
Another worry is that if $\beta_{k}$ and $\beta_{k-1}$ are separated on the two sides of the critical point, we may need a bulk of samples to estimate $Z(\beta_{k-1})/Z(\beta_k)$. 

In addition to the division problem, $Z(\beta_0)/Z(\beta)$ could be an exponentially small number exceeding the data range of computer (e.g. $1.7\times 10^{-308}\sim 1.7\times 10^{308}$ for double type data), and we should store $\ln [Z(\beta_0)/Z(\beta)]$ in the computer. However, simply applying the logarithm to Eq.~(\ref{eq_appx:divide}) and converting the multiplication to addition may lead to precision loss in computer calculations if $\ln [Z(\beta_{k-1})/Z(\beta_k)]$ and $\ln [Z(\beta_{k})/Z(\beta_{k+1})]$ differ by several orders of magnitude.

To address the two problems above, we present the \emph{automatic/pseudo-automatic annealing scheme}.
The automatic annealing scheme starts by fixing $Z(\beta_{k-1})/Z(\beta_k)=\epsilon<1$ for all $k$. We will talk about the choice of $\epsilon$ later, and first assume we have already selected some $\epsilon$. 

Recall that in SSE, the energy estimator is $E(\beta_{k})=-{\langle n\rangle_{\beta_{k}}}/{\beta_{k}}$,
where $n_{k}\equiv \langle n\rangle_{\beta_{k}}$ denotes the average number of non-identity operators at temperature $\beta_{k}$, which itself is also an observable that can be estimated in SSE simulations. After we have achieved an estimation $\tilde{n}_{k}$ of $n_{k}$ from simulations (this is exactly same with measuring energy by counting the number of non-identity operators in each SSE configuration), we can bring the result into $Z(\beta_{k-1})/Z(\beta_k)$ to have
\begin{equation}\label{eq:rough}
    \epsilon = \frac{Z(\beta_{k-1})}{Z(\beta_{k})}=\bigg\langle \bigg(\frac{\beta_{k-1}}{\beta_{k}}\bigg)^n\bigg\rangle_{\beta_{k}} 
    \approx  \bigg(\frac{\beta_{k-1}}{\beta_{k}}\bigg)^{\tilde{n}_{k}}
\end{equation}
Attention that Eq.~(\ref{eq:rough}) is just an approximation, but it gives us a way to divide $[\beta_0,\beta]$. 

Imagine that we start from the $\beta\equiv\beta_m$, which is the right end point of $[\beta_0,\beta]$, then using Eq.~(\ref{eq:rough}), we can determine the position (value) of $\beta_{m-1}$, which is 
\begin{equation}
    \beta_{m-1} = \epsilon^{1/\tilde{n}_k}\beta_m
\end{equation}
Similarly, we can obtain $\beta_{m-2},\beta_{m-3},\cdots$, and it stops when we encounter some $\beta_{j}$ such that $\beta_{j-1}<\beta_0$. Then $\beta_j$ is set to be $\beta_1$ and the division procedure is over. Notice that each $\beta_{k-1}$ must rely on the knowledge of $\beta_{k}$, and each $\tilde{n}_k$ is obtained from the corresponding SSE simulations. 
This division procedure is similar to a standard (thermal or quantum) annealing process, in which we start from an initial parameter point, and step-by-step change it to a target parameter point. Here we go from $\beta\equiv \beta_m$ to $\beta_0$. 
Of course one can inversely go from $\beta_0$ to $\beta$ by considering $Z(\beta)/Z(\beta_0)$ rather than $Z(\beta_0)/Z(\beta)$. 

About the choice of $\epsilon$, since we have to estimate $Z(\beta_{k-1})/Z(\beta_k)$, we hope this value smaller than $1$ but close to $1$ so that the distribution at $\beta_{k-1}$ and $\beta_k$ would have big overlap to make the simulations more efficient.
Just as we perform some common MC simulations to estimate some observable, the MC steps and the number of bins required for an accurate estimation should be tried. Here in our algorithm, it is similar actually. After we fix some $\epsilon$, we have to determine the required MC steps and bins via some trial simulations. We can expect that if we take a very small $\epsilon$, the MC steps required for an accurate estimation would be large. However, no matter what $\epsilon$ we choose, in principle, as long as we have enough MC steps and bins, the estimation of $Z(\beta_{k-1})/Z(\beta_k)$ can be accurate. 
If $\epsilon$ is very close to $1$, we would also have some problems because the number of divided segments $m$ would be very large. Therefore there is some tradeoff about the choice of $\epsilon$.

\subsection{Pseudo-automatic annealing process}\label{subeq:pseudo}
The ``automatic annealing" scheme mentioned above can not be parallelized since we have to achieve each $\tilde{n}_k$ every time from the simulations so that we can go to the next point. Therefore, practically, we give a priori estimate of the value of $\tilde{n}_k$ without doing the simulations. This is what we call the pseudo-automatic annealing scheme used in our paper.

Recall that energy is an extensive quantity, then it should satisfy $E\sim L^{d}$ for a $d$-dimensional system. 
In SSE, we then have $\langle n\rangle_{\beta_k}\sim \beta_kL^d$. Here $\beta_k$, $L$ and $d$ are given, therefore to obtain $\langle n\rangle_{\beta_k}$, we need to know some prefactor $\Lambda_k$ such that $\langle n\rangle_{\beta_k}\simeq \Lambda_k\beta_kL^d\approx \tilde{n}_k$.
It might be questioned that not even we have to decide the value of $\epsilon$, but we have another hyperparameter $\Lambda_k$ to determine. In fact, we still have one since now Eq.~(\ref{eq:rough}) becomes
\begin{equation}\label{eq:require}
    \epsilon \approx \bigg(\frac{\beta_{k-1}}{\beta_k}\bigg)^{\Lambda_k\beta_kL^d}
\end{equation}
If we substitute the real $\Lambda_k$ with some incorrect $\tilde{\Lambda}_k$ we manually set, this just effectively changes the value of $Z(\beta_{k-1})/Z(\beta_k)$ from $\epsilon$ to some $\epsilon_{\text{eff}}$.
Therefore we can fix $\Lambda_k$ to be some value such as $L$ in our simulations for all $k$ for convenience, then the problem reduces to choosing a good $\epsilon$.
The uniform choice of $\Lambda_k\equiv \Lambda$ would make different $Z(\beta_{k-1})/Z(\beta_k)$ slightly deviated from our desired ratio $\epsilon_{\text{eff}}$, but it is okay since we always have to determine the MC steps and bins for estimation. In this aspect, there is no different no matter for $\epsilon_{\text{eff}}$ (with small deviations at each point) or the $\epsilon$ in the ``automatic annealing" scheme in the last subsection.

\begin{figure}[ht!] \centering
    \includegraphics[width=8cm]{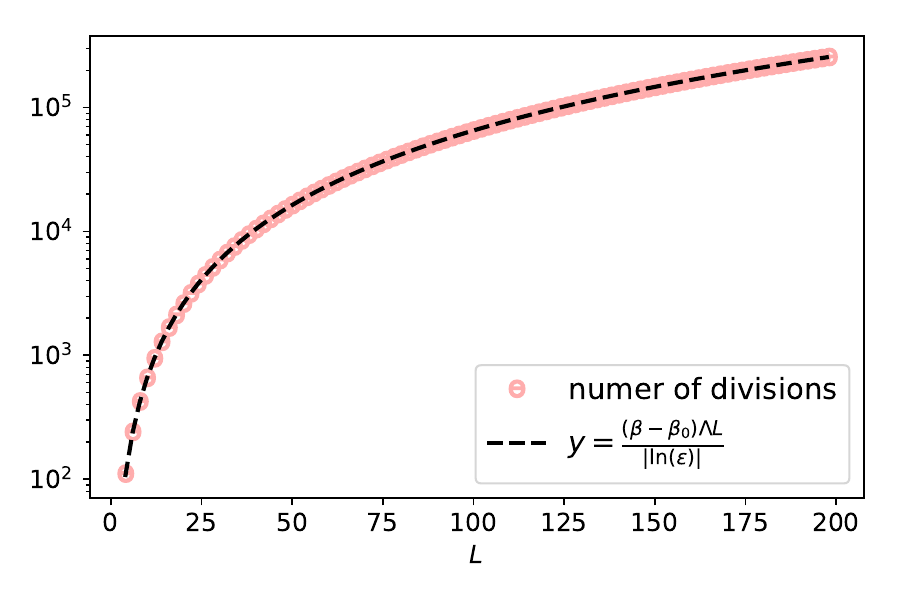}
    \caption{ 
      The number of divisions in the T-Re-An method of first example (see Sec.~\ref{subsec:eg1}), where we take $\epsilon=0.01$, $\beta=30$ and $\Lambda=L$. 
        }
        \label{fig:division}
    % ---------------------------------------------------
\end{figure}

Next we show that this pseudo-automatic annealing scheme actually requires the number of divisions scales proportional to the system size and the inverse temperature for large systems. For the 1D system as an example,
\begin{equation}\label{eq:complexi}
    \beta_{k-1}=\epsilon^{\frac{1}{\Lambda\beta_k L}}\beta_k\approx\bigg(1+\frac{\ln \epsilon}{\Lambda\beta_k L}\bigg)\beta_k
\end{equation}
or
\begin{equation}\label{eqap:divi}
    \beta_{k}-\beta_{k-1}=-\frac{\ln\epsilon}{\Lambda L}
\end{equation}
then the number of divisions is around $(\beta-\beta_0)\Lambda L/|\ln\epsilon|$. Notice that we have ignored the case that $\beta_k\to 0$ in Eq.(\ref{eq:complexi}). This is because when $\beta\to0$, we have $\epsilon^{1/(\Lambda\beta_k L)}\to 0$ ($\epsilon<1$), then the division becomes sparser and has little contributions to the total number of divisions. Another thing worthy of attention is that when $\beta\to0$, the small value of $\epsilon^{1/(\Lambda\beta_k L)}\to 0$ would also exceed the range of a number saved in the computer. Therefore we manually set a lower bound for $\alpha_k$ (e.g. $\min{\alpha_k}=\epsilon$ in our simulation). The existence of this lower bound makes the division in the region that $\beta\to 0$) slightly denser than that Eq.~(\ref{eq:require}) requires. Since most $Z(\beta_{k-1})/Z(\beta_{k})$ still at the same order of magnitude, and the more divisions when $\beta\to 0$ also only constitutes a very small proportion. This affects little to the precision as well as the number of divisions calculated in Eq.~(\ref{eqap:divi}). 
In Fig.~\ref{fig:division}, we plot the number of divisions in the T-Re-An method of first example (see Sec.~\ref{subsec:eg1}) in our paper for reference, where we take $\Lambda=L$, and compare it with the estimation number $(\beta-\beta_0)\Lambda L / |\ln(\epsilon)|$.

% ---------------------------------------------------------
% ---------------------------------------------------------
% ---------------------------------------------------------
% ---------------------------------------------------------
% ---------------------------------------------------------
\section{Examples}\label{sec:examples}
\subsection{1D spin-1/2 Heisenberg chain}\label{subsec:eg1}
As the first example, we consider an 1D spin-1/2 antiferromagnetic (AFM) Heisenberg chain with periodic boundary condition (PBC) with $\beta=30$ and length $L$. We use both the T-Re-An method and the Q-Re-An method to calculate the PF. For the quantum version, we write 
\begin{equation}\label{eq:1d}
    H(s)=s\bigg[\sum_i\mathbf{S}_{i}\cdot \mathbf{S}_{i+1} 
    \bigg] + (1-s) 
    \bigg[ \sum_{i\in\text{even}}\mathbf{S}_{i}\cdot \mathbf{S}_{i+1}\bigg]
\end{equation}
where $H_0=\sum_i\mathbf{S}_{i}\cdot \mathbf{S}_{i+1} $ and $H_1=\sum_{i\in\text{even}}\vec{S}_i\cdot\vec{S}_{i+1}$. 
Apparently, the unique ground state of $H_1$ is a tensor product of $L/2$ singlets, each carrying energy $-3/4$ ($\hbar=1$). 
The estimator for $Z(s')/Z(s'')$ is $\langle (s'/s'')^{n_{\text{odd}}} \rangle_{s''}$ in this model (see the derivation in Appx.~\ref{appdx:estimators}), with $n_{\text{odd}}$ denoting the amount of non-identity operators acting on all the bonds with odd indices within the framework of SSE.

\begin{figure}[ht!] \centering
    \includegraphics[width=9cm]{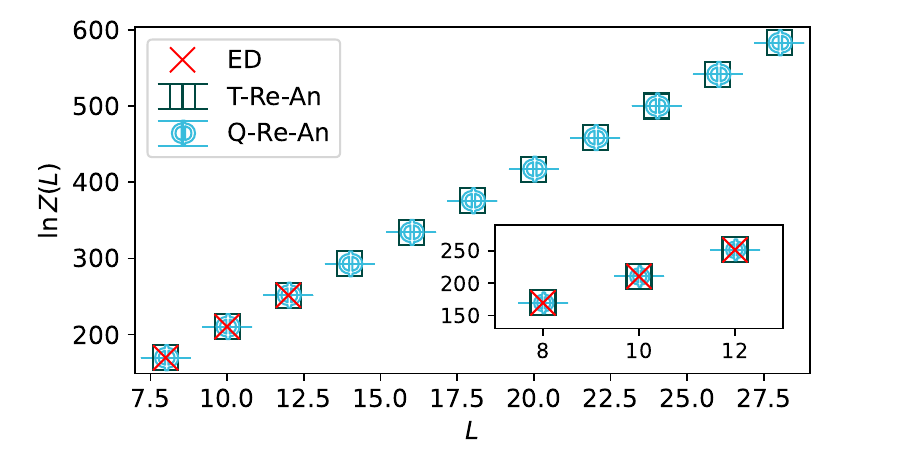}
    \caption{ 
       For $\beta=30$, the values of $\ln Z$ as a function of the chain length $L$, where both T-Re-An method and Q-Re-An method have a perfect match with ED.
       For example, for $L=12$, the results from the two methods are 
        251.625(2) and 251.624(2), respectively, and the exact result is  
       251.62283 from ED calculations. Attention that here in practical SSE simulations, the Hamiltonian has a energy shift of $-1/4$ for each two-body term.
        }
        \label{fig:1d}
    % ---------------------------------------------------
\end{figure}

%In the SSE simulations, we usually add a constant $-1/4$ to each term in Eq.~(\ref{eq:1d}), and the total energy of $H_1$ would be $-L/2$.
%We set the %n=\beta L^2$ in Eq. (\ref{eq:pre}) and 
%$\epsilon=10^{-2}$ in Eq. \ref{eq:pre}. 
%For the system size we simulated, the $\beta$ is large enough for checking the effectiveness of our method in low temperature. 

For the pseudo-annealing scheme, we adopt a similar procedure as in Sec.~\ref{subeq:pseudo}. For $Z(s_{k-1})/Z(s_k)=\langle (s_{k-1}/s_k)^{n_{k,\text{odd}}} \rangle_{s_k}$, we can expect that when $s\to 0$, the number of the operators of the bonds with odd indices goes to zero. Therefore we can take $n_{k,\text{odd}}\approx \tilde{n}_{k,\text{odd}}=\Gamma_k \beta s_kL$ ($\Gamma_k \beta s_k L^2$) for 1D (2D) system, and we find this hypothetical linear relation on $s$ is convenient and has sufficiently good performance just as the first example shows. The number of divisions in this case similarly scales with $\mathcal{O}(\Gamma \beta L)$ for some uniform $\Gamma_k=\Gamma$ if in 1D systems.

As it is shown in Fig. \ref{fig:1d}, both the results of T-Re-An and Q-Re-An method can match that of the exact diagonalization (ED) method exceptionally well for small lengths, which verify the validity of our methods.
%The error bar is about 0.002 while the value is 251.624, and the relative error is smaller than $10^{-7}$. 
The results are also consistent with the expectation that $\ln Z \propto L+o(L)$, where $o(L)$ means higher order terms which is negligible compared with $L$. One may feel this example too trivial and easy to be calculated. Therefore we introduce the second example which needs a high-precision to extract the correction of the sub-leading term in entropy.
\begin{figure}[ht!] \centering
    \includegraphics[width=9cm]{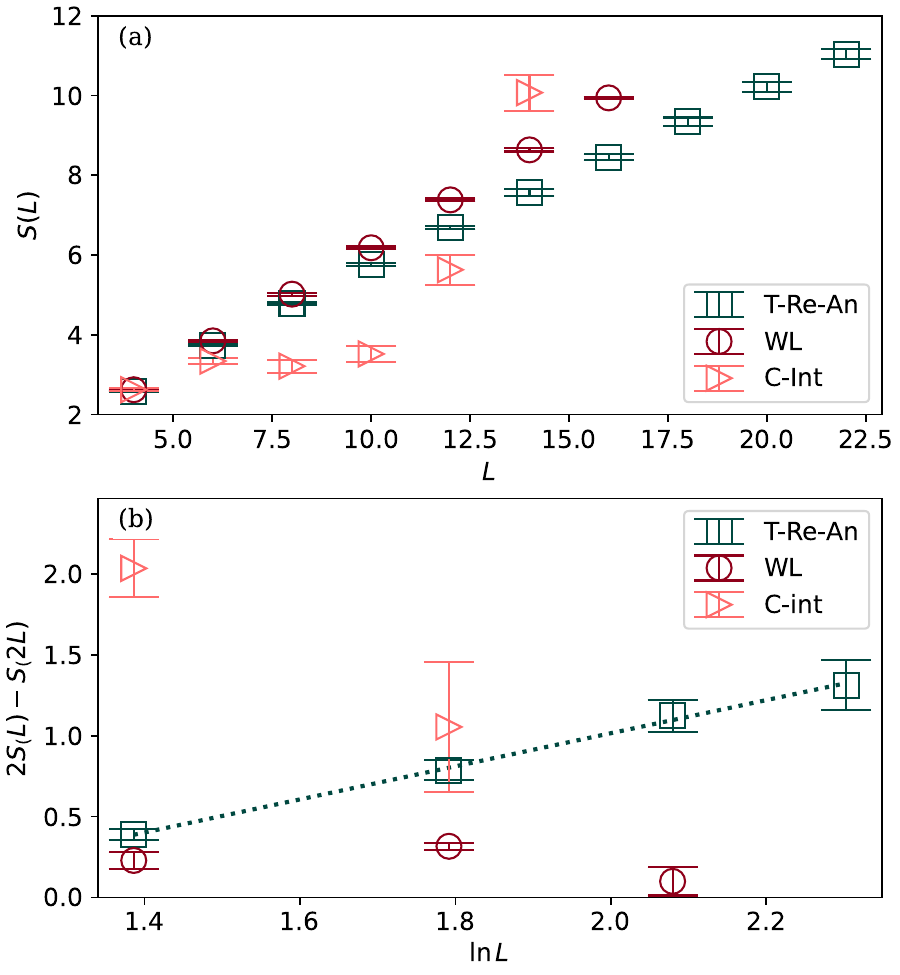}
    % ************************************************
    \caption{ 
       (a) Thermodynamic entropies $S(L)$ as a function of $L$ calculated from different algorithms;
       (b) The $\ln L$ term extracted by calculating $2S(L)-S(2L)$, and the fitting slope (dashed green line) here is 1.00(2).
        }
        \label{fig:2d}
    % ************************************************
\end{figure}

% ---------------------------------------------------------
% ---------------------------------------------------------
% ---------------------------------------------------------
% ---------------------------------------------------------
% ---------------------------------------------------------
\subsection{2D LBW-HM}
To further exploit the advantage of our method, we calculate the thermodynamic entropy of a 2D lattice Bisognano-Wichmann Heisenberg model (LBW-HM)~\cite{PhysRevB.98.134403, Dalmonte2018}, the thermodynamic entropy $S$ of the which at $\beta=1$ provides a good approximation for the Von Neumann entanglement entropy of a 2D PBC square lattice spin-1/2 AFM Heisenberg model under smooth bipartition \cite{PhysRevB.100.155122, Mendes-Santos_2020}. 
A brief review of this theory can be found in Appx.~\ref{appx:bw}.

According to the CFT prediction and previous numerical results~\cite{metlitski2015entanglement,laflorencie2016quantum,zhao2022measuring,deng2023improved}, the coefficient of sub-leading term of the entropy reflects the number of Goldstone modes. Here, the thermal entropy of the LBW-HM at $\beta=1$ should obey $S(L)=aL+b\ln L +c$, and $b=n_G/2$, where $n_G=2$ is the number of Goldstone modes of the 2D spin-1/2 AFM Heisenberg model \cite{1112.5166, laflorencie2016quantum}. 
In this context, coarse result of entropy is disfavored because the the coefficient $b$ in front of the sub-leading term is sensitive to errors. 
We have compared our method with the specific heat integral ($C$-int) and Wang-Landau (WL) algorithm in this example. 

Similar to our method, the $C$-int method also depends on how many intervals we divide for $\beta$. Therefore we take the same division and Monte Carlo steps to compare the performance of these two algorithms. As it shown in Fig.~\ref{fig:2d}, the $C$-int method has great systemic error because the number of division is adequate for our method, but not the $C$-int method, indicating more computational resources needed for it to achieve better accuracy.

For the WL algorithm, 
%on another hand, the allocation of total computational resources should not be predetermined as it hinges on when we can achieve a converged and approximately flat histogram. 
the number of density of states to be flattened in each histogram should be greater than $\beta E(\beta)$, where $E(\beta)$ is the energy at $\beta$, which is also a (roughly) polynomial complexity with the system size \cite{PhysRevLett.90.120201}. However, the main problem of this algorithm is the so-called error saturation, which is intrinsic, preventing us from obtaining results of an arbitrary precision generally and making this algorithm basically biased \cite{fort2015convergence, PhysRevE.78.067701, 10.1063/1.2803061, PhysRevE.72.025701}.
In Fig.~\ref{fig:2d}, we show great error of the results obtained from the vanilla WL samplings \cite{PhysRevLett.90.120201}, and to achieve an arbitrary precision within the Wang-Landau framework is not generally easy.
%This reflects the problem of error saturation, which is intrinsic in this algorithm due to an approximately flat histogram and the empirical refinement parameter~\cite{PhysRevE.78.067701, 10.1063/1.2803061, PhysRevE.72.025701}. 
Another disadvantage of the WL sampling compared with our method is that it cannot be parallelized: the two subroutines for flattening the corresponding histograms before and after an adjustment of the refinement parameter, must be executed sequentially.
In contrast, our unbiased method is able to achieve an arbitrary precision, easy to parallelize, and much more accessible as we only need to measure the observable of $Z(\vec p_{k-1})/Z(\vec p_k)$ in many conventional and mature MC algorithms.

%The WL algorithm, on another hand, suffers from the problem of error saturation, which though can be alleviated by some means like the $t^{-1}$ modification, where $t$ is the Monte Carlo time, the cut-off of the series expansion and the initial value of the refinement parameter $f$ are both hyperparameters to be determined \cite{PhysRevE.78.067701, 10.1063/1.2803061, PhysRevE.72.025701}. Therefore, it is very difficult to achieve results with an arbitrary precision with this method and the convergence time is not generally well-controlled. Besides, in Wang-Landau sampling, the change of $f$ typically obeys some monotonically decreasing function, and the computation subroutines with different $f$'s cannot be parallelized because of the linear logic of the algorithm. In this paper, we consider the vanilla WL algorithm, following Ref. \cite{PhysRevLett.90.120201}, as an example. Fig. \ref{fig:2d}(a) shows that the WL algorithm gives roughly correct result for small sizes while the extraction of entropy in Fig. \ref{fig:2d}(b) completely reveals its unpreciseness, and the $b$ fitted from it is far from the theoretical prediction.

% ---------------------------------------------------------
% ---------------------------------------------------------
% ---------------------------------------------------------
% ---------------------------------------------------------
% ---------------------------------------------------------
\subsection{2D XXZ model}
As the last example, we consider a much more complex model, a 2D spin-1/2 XXZ model with next nearest neighbor interaction, %whose Hamiltonian is 
\begin{align}
    \label{eq:Hj1j2-2}
    H=-\sum_{\langle i,j\rangle}[2S^z_iS^z_{j}+S^x_iS^x_j+S^y_iS^y_j]\nonumber\\
-0.2\sum_{\langle\langle i,j\rangle\rangle}[2S^z_iS^z_{j}+S^x_iS^x_j+S^y_iS^y_j].
\end{align}
where $\langle i,j\rangle$ and $\langle\langle i,j\rangle\rangle$ denotes the nearest and the next nearest neighbors respectively. The lattice we simulated is a square lattice with PBC. This model has a finite temperature 2D Ising phase transition which can be captured by the second or higher order derivative of the free energy $F=-T \ln Z$.

\begin{figure}[ht!] \centering
    \includegraphics[width=9cm]{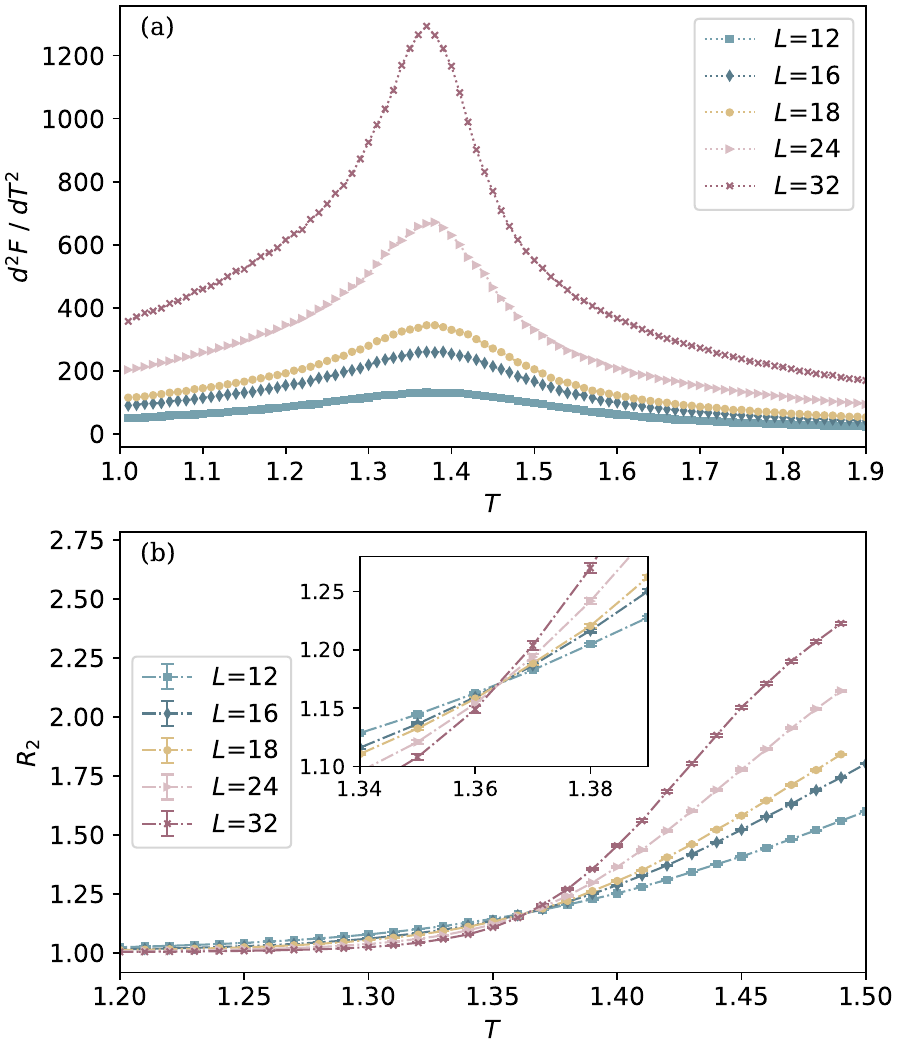}
    \caption{(a) The second order derivative of the free energy, $\partial^2 F/ \partial T^2$, changes with the temperature $T$ in different system size. The peak diverges more when the size becomes larger, which probes the phase transition point here; (b) The Binder ratios in this model at different sizes intersecting at a point which is highly consistent with the peak obtained by Re-An algorithm.
       %\ym{to be edited...}
        }
        \label{fig:ddf}
    % ---------------------------------------------------
\end{figure}

Here we use equal division of $T=1/\beta=0.01$, which is convenient for further numerical differentiation. 
We found that with step size $\Delta T=0.01$ and $20000$ MC samples for each bin, a sufficiently smooth curve can be obtained.
Because a constant factor of the PF does not affect the (higher order) derivative of $F$, we choose $Z(T^*=1.95)$ as the reference point rather than some known point. The PF we calculated is thus $Z(T)/Z(T^*)$ where $Z(T^*)$ is an unknown constant. Then we indeed observe a divergent peak %$\sim T=1.38(1)$ 
of the second order derivative of the free energy as shown in Fig.~\ref{fig:ddf}(a), which accurately probes the phase transition point, matching with the intersection point of Binder ratios $R_2=\langle M^4\rangle/\langle M^2\rangle^2$ of different sizes shown in Fig.~\ref{fig:ddf}(b), where $M$ denotes the magnetization.

This example also inspires us that connecting to a known point is not necessary when we just need the second or higher order derivative of free energy. The data also demonstrate the high-precision of our method, where the second order derivative is still smooth and precise.

% ---------------------------------------------------------
% ---------------------------------------------------------
% ---------------------------------------------------------
% ---------------------------------------------------------
% ---------------------------------------------------------
\section{Summary and discussions}\label{eq:summary}
We propose an unbiased algorithm within the Monte Carlo framework, which has low-technical barrier and no systemic error. The method can be easily implemented for parallel computation and can obtain the value of partition function, free energy and thermal entropy with high precision. 
The method has two keynotes: (1) Finding a parameter path in whole space (either temperature or external parameter) connecting a known point and the to-be-solved one. If we only want to calculate the (high order) derivative of the PF, even fixing an unknown point as the reference is enough; (2) Annealing along the path to calculate the final result through a propagation formula.

We show the advantage of our method via calculating the thermal entropy of 2D LBW-HM and successfully extract precise coefficient of the sub-leading term ($\ln L$), which is consistent with the CFT prediction and catch the correct number of Goldstone modes. In comparison, the $C$-int %which has the same Monte Carlo time with our method 
and the WL sampling both have worse performance. In a 2D XXZ model, the numerical second derivative of the Re-An data is still smooth and precise, which accurately probes the phase transition point.
Our method makes the efficient and high-precision calculation of entropy and other related physical quantities accessible, which can reveal some universal information, such as CFT, criticality and symmetry breaking.

The final thing worth of emphasizing is that the reweight-annealing method discussed here is a general paradigm, not limited in computing the standard partition function, but can be extended to any generalized partition function as long as the two keynotes can be satisfied. One representative example has been shown in \cite{wz24entropy}, where the authors developed an algorithm to compute the R\'enyi entropy using this reweight-annealing framework.

% ---------------------------------------------------------
% ---------------------------------------------------------
% ---------------------------------------------------------
% ---------------------------------------------------------
% ---------------------------------------------------------
\section*{Acknowledgement}
We acknowledge the start-up funding of Westlake University. The authors also acknowledge Beijing PARATERA Tech Co.,Ltd.(\url{https://www.paratera.com/}) for providing HPC resources that have contributed to the research results reported within this paper. CC and ZY thank the support of the National Natural Science Foundation of China (grant no. 12247101). NM thanks the support of the National Natural Science Foundation of China (grant no. 12004020). GP thanks the Würzburg-Dresden Cluster
of Excellence on Complexity and Topology in Quantum
Matter - ct.qmat (EXC 2147, Project No. 390858490).

\textit{Note: After we finished this work, we found a related work posted on arXiv two weeks ago~\cite{dai2024residual}.}

% ---------------------------------------------------------
% ---------------------------------------------------------
% ---------------------------------------------------------
% ---------------------------------------------------------
% ---------------------------------------------------------
\appendix
\section{Estimators in thermal and quantum reweight-annealing under the SSE framework}\label{appdx:estimators}
We first review the reweighting trick in a general MC method with the inverse temperature $\beta$ for illustrations (the thermal case), then we will return to the SSE method, which is just one way of transforming the quantum degrees of freedom to classical ones in the partition function. 

Consider a partition function $Z(\beta)=\sum_C W(\beta; C)$, where the weights $\{W(\beta; C)\}$ are non-negative real numbers (such that we can perform MC simulations) and $C$ denotes each  configuration. 
For the expectation value of some diagonal operator $O$ at the inverse temperature $\beta''$, 
This can be achieved as follows:
\begin{equation}
    \begin{split}
        \langle O\rangle_{\beta''}  =&\frac{\sum_CW(\beta''; C)O(C)}{Z(\beta'')} \\
    \end{split}
\end{equation}
where $O(C)$ is the value of the operator $O$ under the configuration $C$. In our paper, a specific form of the operator $O=[W(\beta')/W(\beta'')]$ is discussed, then we have
\begin{equation}\label{eq:estimate}
\begin{split}
    \bigg\langle \frac{W(\beta')}{W(\beta'')}\bigg\rangle_{\beta''} 
    =& \frac{1}{Z(\beta'')}\sum_CW(\beta''; C)\frac{W(\beta'; C)}{W(\beta''; C)}\\
    =&\frac{\sum_CW(\beta'; C)}{Z(\beta'')} \\
    =&\frac{Z(\beta')}{Z(\beta'')}
\end{split}
\end{equation}

Eq.~(\ref{eq:estimate}) is the key relation in this paper, in which  we sample the configurations with probabilities $p(\beta''; C)=W(\beta''; C)/Z(\beta'')$ at inverse temperature $\beta''$, but recalculating the weights of the sampled configuration $C$ at another inverse temperature $\beta'$, i.e., the $W(\beta'; C)$. 
What Eq.~(\ref{eq:estimate}) indicates is that through averaging the weight ratio of each configuration, the ratio of partition functions at different inverse temperature $Z(\beta')/Z(\beta'')$ can be obtained.

Now we turn to SSE, the demonstrative QMC method used in this paper~\cite{Sandvik1999,Sandvik2010}. 
Apply the Taylor expansion, we have 
\begin{equation}
    Z(\beta)= \text{tr}\bigg\{ \sum_n\frac{\beta^n}{n!}(-H)^n \bigg\} = \sum_n\frac{\beta^n}{n!}\text{tr}\{(-H)^n\}
\end{equation}
Then we write $H=-\sum_{a,b}H_{a,b}$, where $H_{a,b}$ are some non-identity local operator with $a$ denoting the type of it (diagonal or off-diagonal under the simulating basis) and $b$ denotes the index of the space degree (bond or site) that the local operator is acting on. Hence
\begin{equation}
    (-H)^n = \sum_{S_n}\prod_{p=1}^nH_{a_p,b_p}
\end{equation}
where $S_n$ denotes that we sum over all possible operator strings, and each of them is a product of $n$ local operators. Then we have 
\begin{equation}\label{eq2}
    Z(\beta)=\sum_n\sum_{S_n}\frac{\beta^n}{n!}\text{tr}\bigg\{\prod_{p=1}^n H_{a_p,b_p}  \bigg\}
\end{equation}

The important thing here is that because the product of $n$ local operators are interpreted as an operator string, therefore each string $S_n$ has exactly $n$ non-identity operators. In other words, each $S_n$ has the length $n$.

Replacing the trace operation with a set of complete basis $\{\alpha_0\}$, we have
\begin{equation}
Z(\beta)=\sum_n\sum_{S_n}\sum_{\alpha_0}\frac{\beta^n}{n!}\bigg\langle\alpha_0\bigg|\prod_{p=1}^n H_{a_p,b_p}  \bigg|\alpha_0\bigg\rangle
\end{equation}

Since practically we cannot simulate $n$ to infinity, we introduce a large cut-off $M$ to truncate the series. For those $n<M$, after inserting some identity operators (also some special $H_{a,b}$ by using $a$ to denote identity, diagonal and off-diagonal types), we can make all operator strings have the same length $M$. Besides, we introduce $(M-1)$ extra sets of complete basis, and finally we have 
\begin{equation}\label{eq:sse_last}
    Z(\beta)=\sum_{S_M}\sum_{\alpha}\frac{\beta^n(M-n)!}{M!}\prod_{p=1}^M\langle \alpha_{p-1}|H_{a_p,b_p}|\alpha_p\rangle 
\end{equation}
where we have identify $\alpha_M\equiv \alpha_0$ and $\alpha\equiv \alpha_0,\cdots,\alpha_{M-1}$. 

The weight in Eq.(\ref{eq:sse_last}) is 
\begin{equation}
    W(\beta;S_M,\alpha)= \frac{\beta^n(M-n)!}{M!}\prod_{p=1}^M\langle \alpha_{p-1}|H_{a_p,b_p}|\alpha_p\rangle 
\end{equation}
and the classical configurations are given by both the states $\alpha$ and the operator string $S_M$.
By replacing the weights $W(\beta'; C)$ and $W(\beta''; C)$ in Eq.~(\ref{eq:estimate}) 
with $W(\beta';S_M,\alpha)$ and $W(\beta'';S_M,\alpha)$, we have
\begin{equation}\label{eq:ssss}
    \frac{Z(\beta')}{Z(\beta'')}=\bigg\langle \bigg(\frac{\beta'}{\beta''}\bigg)^n\bigg\rangle_{\beta''}
\end{equation}
for $n$ to denote the number of non-identity operators.

Similar discussions can be applied to the quantum case. For  $H(s)=sH_0+(1-s)H_1=H_1+s(H_0-H_1)$, we usually have $Z(s')/Z(s'')=\langle (s'/s'')^{n_s}\rangle_{s''}$ within the framework of SSE, where $n_s$ is the number of non-identity operators related to $(H_0-H_1)$. For instance, if
\begin{equation}
    \begin{split}
    H(s)=&s\bigg[\sum_i\mathbf{S}_{i}\cdot \mathbf{S}_{i+1} 
    \bigg] + (1-s) 
    \bigg[ \sum_{i\in\text{even}}\mathbf{S}_{i}\cdot \mathbf{S}_{i+1}\bigg]\\
    =&\sum_{i\in\text{even}}\mathbf{S}_{i}\cdot \mathbf{S}_{i+1} + s\bigg[
    \sum_{i\in\text{odd}}\mathbf{S}_{i}\cdot \mathbf{S}_{i+1}
    \bigg]
    \end{split}
\end{equation}
where $H_0=\sum_i\mathbf{S}_{i}\cdot \mathbf{S}_{i+1} $ and $H_1=\sum_{i\in\text{even}}\vec{S}_i\cdot\vec{S}_{i+1}$, then we can obtain
\begin{equation}\label{eq_appx:rss}
\begin{split}
	\frac{Z(s')}{Z(s'')}=&\frac{1}{Z(s'')}\text{tr}\{e^{-\beta H(s')}\}\\
	=&\frac{1}{Z(s'')}\sum_{\alpha,S_M}\bigg(\frac{s'}{2}\bigg)^{n_{\text{odd}}}\bigg(\frac{1}{2}\bigg)^{n_{\text{even}}}\tilde{S}_{\alpha,M} \\
	=&\frac{1}{Z(s'')}\sum_{\alpha,S_M}
	\bigg(\frac{s'}{s''}\bigg)^{n_{\text{odd}}}
	\bigg(\frac{s''}{2}\bigg)^{n_{\text{odd}}}\bigg(\frac{1}{2}\bigg)^{n_{\text{even}}}\tilde{S}_{\alpha,M} \\
	=&\bigg\langle \bigg(\frac{s'}{s''}\bigg)^{n_{\text{odd}}}\bigg\rangle_{s''}\\
\end{split}
\end{equation}
where $n_{\text{odd}}$ ($n_{\text{even}}$) denotes the amounts of non-identity operators acting on the bonds with odd (even) indices in SSE simulations, and  
\begin{equation}
    \tilde{S}_{\alpha,M}=\frac{\beta^n(M-n)!}{M!}\prod_{p=1}^M\langle\alpha_{p-1}|H_{a_p,b_p}|\alpha_p\rangle.
\end{equation}

% ---------------------------------------------------------
% ---------------------------------------------------------
% ---------------------------------------------------------
% ---------------------------------------------------------
% ---------------------------------------------------------
\section{Computing entanglement entropy}\label{appx:bw}
One important application of our method is to calculate the entanglement entropy in conjunction with the Bisognano-Wichmann  theorem, which offers an excellently approximated functional form to the entanglement Hamiltonian in systems exhibiting translational symmetry~\cite{PhysRevB.98.134403, Dalmonte2018}. We briefly introduce the functional form of 1D/2D system with periodic boundary condition here. 

\begin{figure}[ht!] \centering
    \includegraphics[width=8cm]{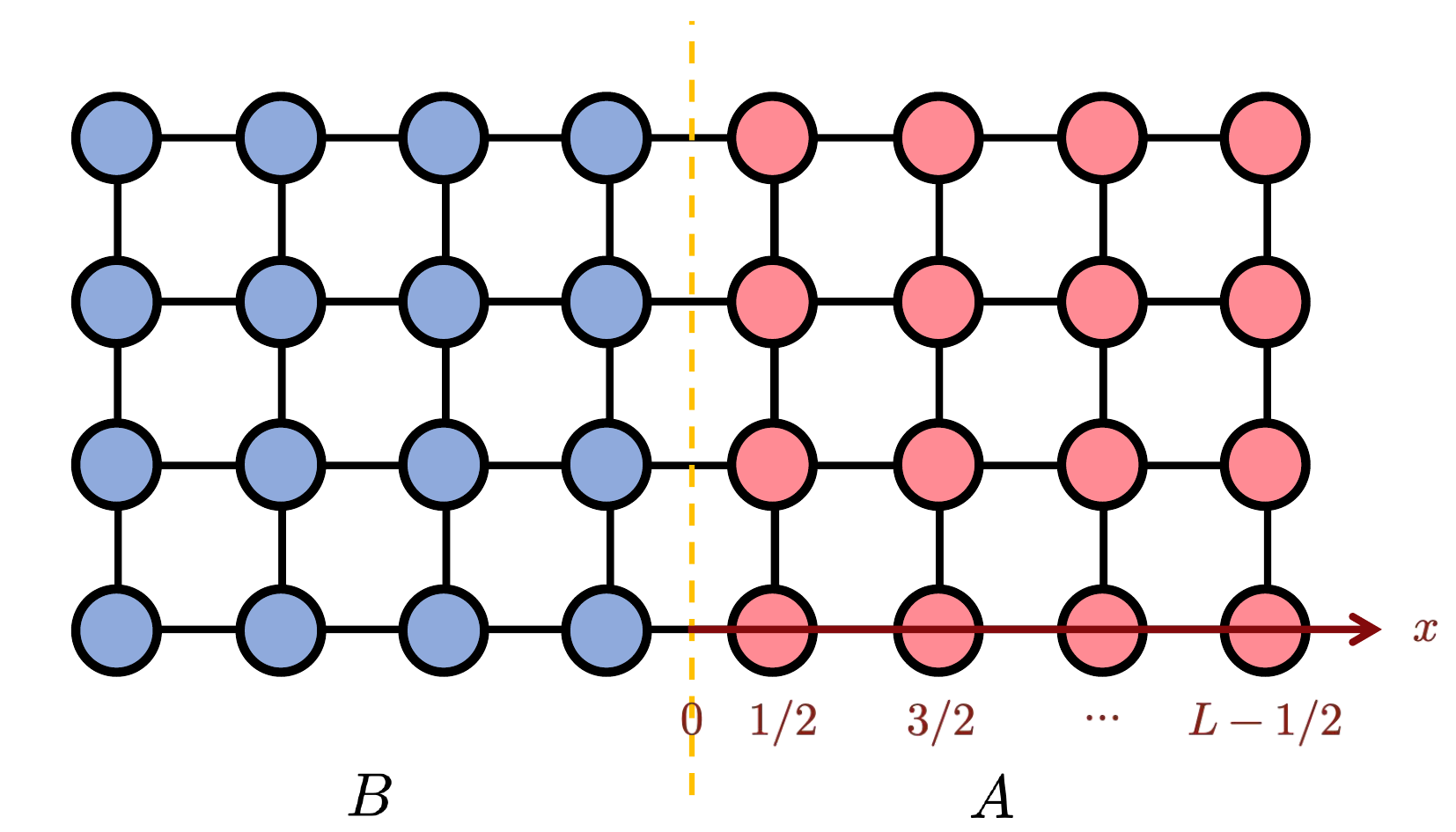}
    \caption{ 
       Illustration of the LBW Hamiltonian with an example of a 2D square lattice with size $2L\times L$ under half bipartition. The $d$ in Eq.~(\ref{eq_appx:bw_coff}) denotes the distance from the corresponding site/bond to the boundary (the orange dashed line). The spacing of lattice is taken to be $1$.
        }
        \label{fig:bw_sch}
    % ---------------------------------------------------
\end{figure}

\begin{figure}[ht!] \centering
    \includegraphics[width=9cm]{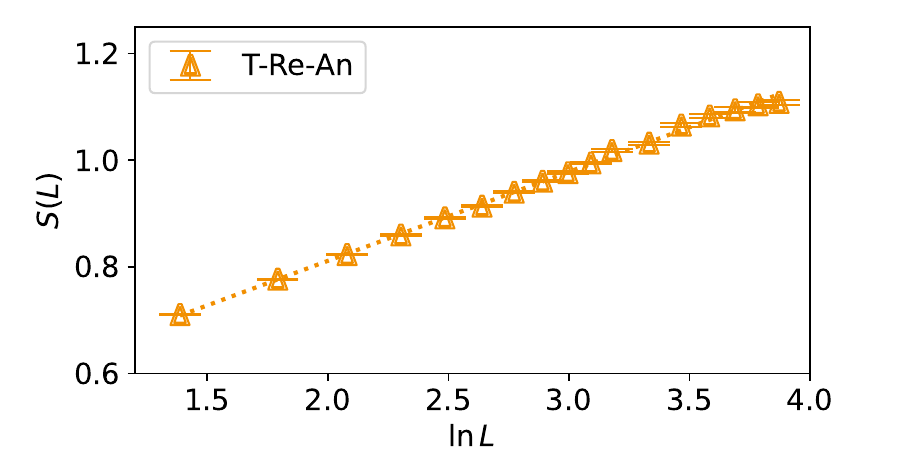}
    \caption{ 
       For the 1D TFIM model at the critical point under PBC, the Von Neumann entropy obtained via calculating the thermal entropy of the corresponding BW-TFIM model at $\beta=1$. The central charge extracted from our Re-An simulations is $c=0.501(4)$, which is consistent with the theoretical result $0.5$.
        }
        \label{fig:1d_tfim}
    % ---------------------------------------------------
\end{figure}

As it shown in Fig.~\ref{fig:bw_sch}, for an 1D/2D system under half bipartition (cornerless) with $A$ and $B$ part, if the Hamiltonian takes the form
\begin{equation}
    \begin{split}
    H=&\Gamma\sum_{x,y,\delta=\pm 1} [h_{(x,y),(x+\delta,y)} + h_{(x,y),(x,y+\delta)}] \\
    &+ \Theta\sum_{x,y}l_{(x,y)}\\
    \end{split}
\end{equation}
with nearest-neighbor coupling $\Gamma$ and on-site coupling $\Theta$, then the corresponding lattice Bisognano-Wichmann (LBW)  Hamiltonian $\tilde{H}_A$ is 
\begin{equation}
\begin{split}
    \tilde{H}_A=&\frac{2\pi}{v}\bigg[ \sum_{x,y,\delta=\pm 1}
    [\Gamma_x h_{(x,y),(x+\delta,y)} + \Gamma_y h_{(x,y),(x,y+\delta)}] \\
    &+ \Theta_x\sum_{x,y}l_{(x,y)}\bigg]
\end{split}
\end{equation}
where $v$ denotes the sound velocity and $\Gamma$, $\Theta$ are modified to be $\Gamma_x,\ \Gamma_y$, and $\Theta_x$. For period boundary condition, we have 
\begin{equation}\label{eq_appx:bw_coff}
\begin{split}
    \Gamma_{\delta=x,y}&=\frac{d(L-d)}{L} \\
    \Theta_x&=\frac{d(L-d)}{L} \\
\end{split}
\end{equation}
which depend on the distance $d$ to the boundary of the corresponding bond/site term (See Fig.~\ref{fig:bw_sch}).

With the LBW Hamiltonian $\tilde{H}_A$, we can define the LBW reduced density matrix $\tilde{\rho}_A$, which is
\begin{equation}
    \tilde{\rho}_A\equiv \frac{e^{-\tilde{H}_A}}{\text{tr}\{e^{-\tilde{H}_A}\}}\approx \rho_A
\end{equation}
where $\rho_A$ is the real reduced density matrix. Therefore we can compute the Von Neumann entropy $S(\rho_A)$ by 
\begin{equation}\label{eq:bwvne}
    S(\rho_A)\approx -\text{tr}\{\tilde{\rho}_A\ln \tilde{\rho}_A\} = \langle \tilde{H}_A\rangle + \ln \tilde{Z}_A
\end{equation}
where $Z_A\equiv \text{tr}\{e^{-\tilde{H}_A}\}$. Attention that Eq.~(\ref{eq:bwvne}) is exactly the thermal entropy of $\tilde{H}_A$ at $\beta=1$. Since it requires the value of $\ln \tilde{Z}_A$, our method can thus be used to calculate the Von Neumann entropy of lattice systems with translational symmetry.

In addition to the LBW-HM model shown in the main body of the paper, we present an additional example of the 1D transverse field Ising model (TFIM), and calculate its Von Neumann entropy at the critical point with the reweight-annealing method. The exact result is $S(L)\propto (c/3)\ln L$, where $c=0.5$ is the central charge, and our method gives $c\approx 0.501(4)$, as it shown in Fig.~\ref{fig:1d_tfim}.

% ---------------------------------------------------------
% ---------------------------------------------------------
% ---------------------------------------------------------
% ---------------------------------------------------------
% ---------------------------------------------------------
\bibliography{ref}

\end{document}